# Methods for Predicting Behavior of Elephant Flows in Data Center Networks

Aymen Hasan Alawadi [1], Maiass Zaher [2], and Sándor Molnár [3]

*Abstract*—Several Traffic Engineering (TE) techniques based on SDN (Software-defined networking) proposed to resolve flow competitions for network resources. However, there is no comprehensive study on the probability distribution of their throughput. Moreover, there is no study on predicting the future of elephant flows. To address these issues, we propose a new stochastic performance evaluation model to estimate the loss rate of two state-of-art flow scheduling algorithms including Equal-cost multi-path routing (ECMP), Hedera besides a flow congestion control algorithm which is Data Center TCP (DCTCP). Although these algorithms have theoretical and practical benefits, their effectiveness has not been statistically investigated and analyzed in conserving the elephant flows. Therefore, we conducted extensive experiments on the fat-tree data center network to examine the efficiency of the algorithms under different network circumstances based on Monte Carlo risk analysis. The results show that Hedera is still risky to be used to handle the elephant flows due to its unstable throughput achieved under stochastic network congestion. On the other hand, DCTCP found suffering under high load scenarios. These outcomes might apply to all data center applications, in particular, the applications that demand high stability and productivity.

*Index Terms*— Elephant flow, SDN, Risk analysis, Value-at-Risk, Flow scheduling, Congestion control.

## I. INTRODUCTION

Nowadays, many enterprises leverage data center fabrics to manage highly-demanded bandwidth applications. Applications like Hadoop [1] and MapReduce [2] rely on hundreds or thousands of servers to provide high availability and scalability; therefore large data is transferred through the data center network to achieve these requirements. However, other types of data center applications such as regular web services are hosted inside the data center as well, due to the guaranteed availability and reliability. Because of these substantial requirements, many data center topologies evolved like hyperx [3], flattened butterfly [4], and fat-tree [5]. On the other hand, many traffic management techniques emerged, like throughput-based forwarding and load balancing [6]. Typically, the applications of data center produce two types of flows which are mice and elephant flows [6]. Mice flows are known as the smallest and shortest-lived TCP flows in the network and more sensitive to the communication delay. Whereas the most massive and long-lived TCP flows, elephant flows, are more affected by the residual link bandwidth [6].

Department of Telecommunications and Media Informatics, Budapest University of Technology and Economics, Hungary, 1117 Budapest, Magyar Tudosok krt. 2.
[1] aymen@tmit.bme.hu, [2] zaher@tmit.bme.hu,
[3] molnar@tmit.bme.hu

The number of elephant flows in data centers is fewer than that of mice flows, but they carry the most, e.g., 80%, of the transferred data [7]. Some applications, like data mining, machine learning, and data analysis [8] [9] generate such flows since they demand intensive data transmission. These flows must be forwarded through appropriate routes following their requirements. Static forwarding techniques like ECMP [10] could yield network congestions where bottlenecks would stem from collides on a specific switch port due to static hashing [11] [12]. Hence, enhancing flow scheduling in data center networks would improve throughput and Flow Completion Time (FCT).

In today's data centers, SDN plays a vital role in network resource allocation, traffic monitoring, and classification [14]. The paradigm has significantly employed by the research community for flow scheduling, and traffic load balancing [15] [16] since the implementation of real-time applications is delicate without adequate resource and traffic management [2]. The standard design of a data center network includes multi-rooted trees that have multiple paths between every pair of hosts [12]. As a result, the challenge is to identify the suitable path for flows according to the current load of the paths and to avoid network congestion. However, most of the existing flow scheduling solutions like Hedera [12] forward both flow types on the same paths; hence, flow competitions and bottlenecks are inevitable [17]. Furthermore, rerouting the elephant flows might yield delay, packet reordering, and retransmission.

In this paper, we evaluate and predict the performance of ECMP, Hedera, and DCTCP. Particularly, we empirically investigate the performance and efficiency of the algorithms to answer the following questions:

1. What is the predicted loss rate of elephant flows using different algorithms?
2. What are the risk factors of implementing these algorithms regarding the elephant flow preserving?
3. How could the FCT and throughput of mice and elephant flows be under different algorithms?

Therefore, our main contributions are:

1. Implementing a wide range of workloads to estimate the probability distribution of the algorithms' performance.
2. Conducting stochastic performance analysis instead of deterministic one to explore the minimum and maximum value of elephant flows loss rate.
3. Predicting the future performance of the different algorithms based on the stochastic evaluation and

demonstrate their impact on data center applications in terms of the expected productivity.

The rest of the paper is organized as follows. In section II, we present related works. We describe the proposed model in section III. In section IV, we describe the simulations, results and, discussions. We finally conclude in Section V.

## II. RELATED WORKS

Liu et al. [18] present a framework to enable adaptive multipath routing of elephant flows in data center networks under changing load conditions; however, this solution employs NOX controller which has some negative effects on the performance. Similar to Mahout [15], it detects elephant flows at end-hosts, but it monitors TCP socket buffer at end-host to mark flows exceed a predefined threshold so that elephant flows are forwarded based on a weighted multipath routing algorithm which results in installing better paths in switches. Besides, like Hedera, mice flows are delivered based on ECMP by default. However, it employs link load as the only metric for rerouting decisions. Devoflow [19] provides a flow control mechanism in data center networks by rerouting elephant flows whose sizes are more significant than 1 MB.

Similarly, authors in [20] employ group feature of OpenFlow to implement a framework for managing the routes in data center networks by checking links load so that the framework distributes flows among different paths to balance the loads. This framework provides no distinguishing between elephant and mice flows, but when the congestion occurs on a link, the framework selects a backup flow with most considerable traffic demand, which means in practice most probably it will be an elephant flow, but it does not provide any measurements about the impact on mice flows. Wang et al. in [21] present TSACO, which detects elephant flows by OpenFlow and sFlow then forwards them according to an adaptive multi-path algorithm and handles mice flows differently. TSACO computes the available bandwidth and delay of paths and splits an elephant flow over multiple paths, which have considerably enough free bandwidth to balance the load whereas it sends mice flows on the remaining computed flows whose delay characteristics are suitable. As a result, TSACO provides better throughput for elephant flows, and shorter delay for mice flows in comparison with ECMP and weighted ECMP.

## III. EXPERIMENTAL METHODOLOGY

In this section, we describe our experimental methodology, including our system setup, network setup, and applications workloads employed in our empirical study.

### A. System setup

*K*-4 fat-tree data center topology was built by using Mininet 2.2.2 SDN emulator installed on Ubuntu 16.04 machine provided with Intel Core i5-8400 CPU 2.80 GHz with 16 GB of RAM.

### B. Flow scheduling algorithms

1. **Hedera**: estimates the demand for elephant flows then reroute them to a path with sufficient bandwidth by installing new flow entries on the switches. Particularly, flows will be forwarded through one of the equal-cost paths by applying a static hashing based technique as in ECMP until they grow beyond the predefined threshold which is 10% of the link capacity [12].
2. **Equal-Cost Multi-Path (ECMP)**: switches are statically configured with several forwarding paths for different subnets. The forwarding is based on the hash value of specific fields of packets header modulo the number of paths for spreading the load across many paths [10].
3. **DCTCP**: employs Explicit Congestion Notification (ECN) to estimate the fraction of bytes that encounter congestion rather than directly detecting that congestion has occurred. Then, DCTCP scales the size of the TCP congestion window accordingly. This method provides low latency and high throughput with shallow-buffered switches where they can be used in large data centers to reduce the capital expenditure. In typical DCTCP deployments, the marking threshold in the switches is set to a deficient value to reduce queueing delay, and a relatively small amount of congestion will cause the marking. During the blockage, DCTCP will use the fraction of marked packets to reduce the size of the congestion window more gradually than that in case of conventional TCP [22].

DCTCP and Hedera algorithms are implemented and tested as SDN applications by using Ryu controller whereas, ECMP is implemented statically in switches.

### C. Collecting and normalizing the data

In this section, we present the conducted experiment to evaluate the results of the proposed evaluation model. In this paper, fat-tree topology is used since it is considered one of the essential topologies for building efficient, scalable, and cost-effective data centers. Fat-tree topology constructed from three main layers of connected switches located in core, aggregate, and edge layers. However, *K*-4 fat-tree data center topology has been built in Mininet with 10 Mbps links for each as shown in Figure 1.

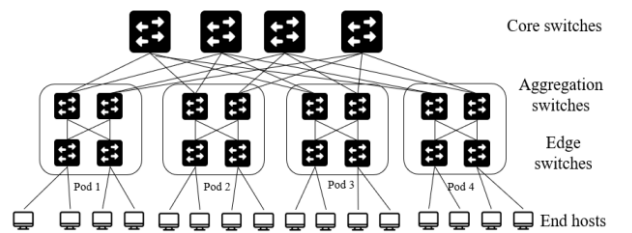

Fig. 1 K-4 fat-tree data center.

The conducted scenarios have two patterns; the first one generates connections that span all topology layers while the second one generates connections span the switches in edge and aggregation layers only as depicted in Figure 2. In these patterns, all of the end hosts in each rack employed to generate the traffic for each of the proposed scenario. To generate the required elephant and mice flows, we employed *iperf* for generating elephant flows, whereas the traffic of mice flows was generated by requesting specific files whose sizes are 10 Kbyte by applying an Apache server repeatedly in a random fashion as reported in [7].

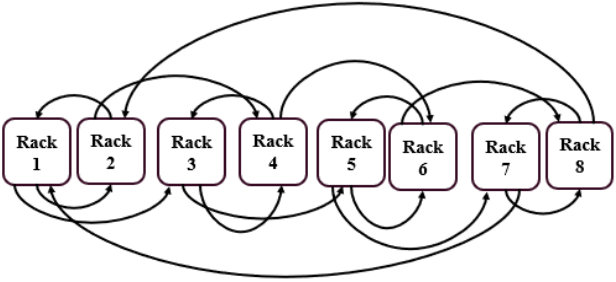

Fig. 2 The traffic pattern.

However, the performance of the proposed model will be evaluated under high load scenarios where the mice flows are synchronized with the elephant flows to introduce congestion in the network. The evaluation process includes three different scenarios with different workloads, a mix of elephant and mice flows, whose time span are varied from 1 to 15 seconds in case of elephant flows to evaluate the investigated algorithms with different sizes of elephant flows. In the first scenario, 1:1 ratio, we generated 120 concurrent connections, so mice and elephant flows have an equal proportion, e.g., 60:60, respectively. In the second scenario, the 1:2 ratio, where we increased the number of the elephant flows to 80, and reduced mice flows to 40. Finally, in the third scenario, 2:1 ratio, where we have 40 elephant flows to 80 mice flows. However, each scenario has been executed twenty-five times, and during each repetition, the throughput has been measured between hosts 1 and 16 by creating a 20 seconds connection using *iperf* to reflect the impact of different algorithms on the throughput of a specific elephant flow, and we built our risk analysis based on it. To obtain the risk factor of the error in throughput measurements, we utilized the arithmetic sample standard deviation. The maximum value of the calculated standard deviation is considered since it indicates a more significant value than the sample mean for the worst-case evaluation.

### D. Goodness of fit

The goodness of fit test performed to find the proper probability distribution functions of the throughputs and errors. Therefore, we adopted EasyFit professional [23], which is a specialized statistical tool to test the collected data. Since the collected data is in the discrete domain, we chose to use the Kolmogorov Smirnov statistic test (KS) as a hypothesis test to assess the distribution of the data [24]. KS test is a non-parametric test mainly used to compare the distance between the empirical data samples and a specific class of well-known reference probability distributions as in equation 1 [25].

$$D_n = \sup_x |F_n(x) - F(x)| \quad (1)$$

Where $F_n$ is the cumulative distribution function of the observed samples in comparison with the reference distribution functions $F$ of an ordered data.

A null hypothesis testing has been performed to accomplish this kind of testing, where $H_0$ is identified when the tested data specify the distribution, and $H_1$ is recognized when the data does not follow the distribution. To come up with the desired distribution, KS assumes a significance level α (0.01, 0.05, etc.) and compares the tested statistics ($D_n$) with some of the critical values of the well-known distribution. The hypothesis of the measured distribution will be discarded if the value of $D_n$ exceeds the critical value at a significant level.

P-value based on the KS test helps to identify the level when the null hypothesis is rejected. This value indicates a threshold for the significant level ($H_0$) to accept all values less than the P-value. For instance, when the P-value = 0.025, the null hypothesis will take all the significance levels less than the P-value, i.e., 0.01 and 0.02, and reject the higher levels [26].

Table 1 shows the results of conducting KS null hypothesis testing on the throughput measurements of the algorithms. The throughput of both Hedera and ECMP followed the Geometric distribution (G-D) based on P-value, and the acceptable critical value was 0.02. However, G-D is recognized as a discrete probability distribution that represents the probability of the success number of independent trials, i.e., Bernoulli trials [27].

TABLE I. KS TEST VALUES FOR THE AVAILABLE THROUGHOUT.

| Algorithm | KS accepted values (critical values) | P-Value | Distribution |
|---|---|---|---|
| Hedera | 0.05 | 0.07077 | Geometric |
| ECMP | 0.02 | 0.03 | Geometric |
| DCTCP | Rejected | 0.008 | - |

We got rejection as a result of DCTCP distribution testing for all of the significance levels, as appeared in Table 1. Therefore, we used another normality test called the Anderson-Darling (AD) test. However, the AD test followed the null hypothesis testing and defined as $A^2$.

$$A^2 = -N - S \quad (2)$$

Where $S$:

$$S = \sum_{i=1}^{N} \frac{(2i-1)}{N} [\ln F(Y_i) + \ln(1 - F(Y_N + 1 - i))] \quad (3)$$

Where $F$ is the cumulative distribution function of the observed samples and $Y_i$ are the ordered data.

The testing shows that the throughput of DCTCP followed G-D with an acceptable critical value equals 0.02. Hence, we utilized probability mass function of G-D to generate samples required for Monte Carlo simulation model by applying equation 4 where Hedera, ECMP, and DCTCP have different probability values.

$$P_r(A) = (1-p)^{r-1} p \quad (4)$$

Where $A$ is the random variable of the throughput, $r$ is the number of failures with $p$ probability.

Besides, we repeated the same procedure, to identify the distribution of measurement errors, i.e., error factors, as shown in Table 2.

TABLE II. KS TEST VALUES FOR ERROR FACTOR.

| Algorithm | KS accepted values (critical values) | P-Value | Distribution |
|---|---|---|---|
| Hedera | 0.05 | 0.86674 | Discrete uniform |
| ECMP | 0.05 | 0.2179 | Negative binomial |
| DCTCP | 0.05 | 0.76964 | Poisson |

In the case of Hedera, the testing of Hedera algorithm showed that it was following the Discrete Uniform distribution (D-U). Therefore, to generate the required samples of the error factor, we used equation 5.

$$P_r(E_{rH}) = \frac{1}{n} \quad (5)$$

Where $E_{rH}$ is the random variable of the error factor for Hedera algorithm and $n$ is the number of samples generated for the error factor.

On the other hand, the error factor of the ECMP algorithm followed Negative binomial distribution (N-B). N-B is a discrete probability distribution mainly describes the number of successes in a series of independent Bernoulli trials until arriving the defined number of non-random failures occurs. Hence, to generate its sample values for error factor, we utilized equation 6.

$$P_r(E_{rE}) = \binom{r+c-1}{c} p^c (1-p)^r \quad (6)$$

Where $E_{rE}$ is the random variable of the error, $r$ is the number of failures with $1-p$ probability, $c$ is the number of success or failure and, $p$ is the probability of success.

Similarly, the error factor for DCTCP followed a Poisson distribution (P-D) function. However, P-D mostly used to express the probability of occurring certain events within the sample space or fixed interval of time [28]. The probability mass function of the P-D, i.e., equation 7, was used to generate its required samples of the error factor.

$$P_r(E_{rD}) = e^{-k} \frac{\lambda^k}{k!} \quad (7)$$

Where $E_{rD}$ is the random variable of the error factor, $\lambda$ is the average number of errors recorded per the whole sample, $e$ is the Euler's number 2.71828, and $k$ is the number actually observed occurrences.

### E. Monte Carlo Simulation

Monte Carlo approach is a technique used to reproduce the stochastic behavior of a system or to assess a set of uncertainty input of a deterministic model. Typically, it is not possible to predict and determine all possible outcomes of a black box system [29]. Hence, the Monte Carlo simulation process utilized to generate multiple predicted scenarios by estimating the probability distribution of the stochastic input parameters. Consequently, this process recurred hundreds or thousands of times to produce possible scenarios or solutions with different probabilities.

However, we address the impact of the algorithms by calculating the value at risk of the elephant flows. For this purpose, we used the generated samples of throughput and error factor as inputs of the Monte Carlo simulation model. Our fundamental equation, i.e., equation 8 that forms Monte Carlo simulation based on simulating various sizes and volumes of elephant flows along with the risk values.

$$Pre\,(V, S, A, E) = B_m = V_i \times (S_j - (A_k + E_l)) \quad (8)$$

Where $B_m$ is the predicted loss rate, $V_i$ is the different volumes of the evaluated elephant flows, $S_j$ is the sizes of the elephant flows, as shown in Table 3, $A_k$ is the available throughput factor, and $E_l$ is the error factor variables.

TABLE III. ELEPHANT FLOW PARAMETERS.

| Elephant flow | Size S | Volume V |
|---|---|---|
| Large | 1.25 MByte | 100 |
| Normal 1 | 0.75 MByte | 85 |
| Normal 2 | 0.5 MByte | 65 |
| Small | 0.12 MByte | 45 |

The assumed values for the size $S$ varies from the maximum bandwidth the physical link can handle, i.e., 10 Mbps, to the minimum elephant flow size, i.e., 10% of link capacity, as defined by Al-Fares et al. [12]. The volume parameter $V$ represents the amount of the flow within a specific path.

## IV. RESULTS AND DISCUSSIONS

The proposed model performed on the algorithms to investigate how they will preserve the elephant flows. The primary expected outcome from this analysis is a histogram represents the probability distribution of the predicted loss rate of elephant flows resulted from employing each algorithm. Therefore, equation 8 repeated one million times. On the upcoming subsections, we will address and compare the results of the investigated algorithms.

### A. Throughput of the elephant flow

In Figure 3, we compare the achieved throughput of the algorithms under different scenarios by tracking the connection between hosts 1 and 16. Furthermore, to measure the stability of each algorithm, we calculated the second central moment, e.g., error variance. Results show that Hedera achieved the highest variance, 25.5, in comparison with ECMP, 21.74, while DCTCP had 17.17.

### B. Loss rate distribution

Monte Carlo simulation provided the whole estimation for the tested data, as shown in Table 4. However, it is clear that the DCTCP algorithm achieved the worst loss rate due to the fact that DCTCP does not provide any special handling for elephant flows. Furthermore, Hedera and ECMP have layer-4 flow control mechanisms and scheduling capabilities as well.

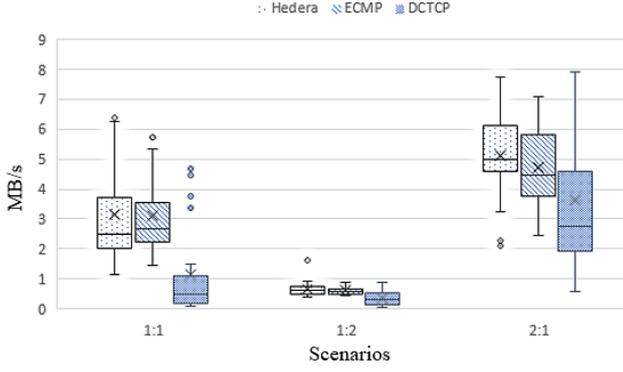

Fig. 3 Throughput achievement of the different scenarios.

TABLE IV. DISTRIBUTION STATISTICS OF THE TESTED ALGORITHMS.

| Algorithm | Loss rate |
|---|---|
| Hedera | 64% |
| ECMP | 68% |
| DCTCP | 77% |

## C. Distribution shape analysis

To implement the value at risk (VaR) analysis for the obtained result, we should present the histogram of loss rate for each algorithm to figure out the variations of the values. Figures 4, 5, and 6 depict the histograms for Hedera, ECMP, and DCTCP, respectively. The current histograms do not follow a particular type of known probability distributions, but we can indicate that they have a heavy left-hand tail and unsteady proceed to the long right-hand tail. Nevertheless, the yielded histograms may plot almost the same behavior regarding the shape, since the input values ($S$ and $V$) are the same. Considering the first raw moment of the mean value, blue line, and median value, red line, for such samples may not present the exact expected value of the loss rate [30] since the prediction depends on the merging of the risk values of different samples. Consequently, since we have the sufficient number of samples and for better understanding of the behavior of the loss rate, common distribution shape measurements were calculated, like skewness and kurtosis, as shown in Table 5.

Skewness is the third central moment, and it used for measuring the symmetry of the distribution, and it has two values; positive and negative. The positive value, i.e., right skew, indicates that the mean value is higher than the median value, while the negative value, left skew, suggests the opposite.

Equation 9 describes the skewness degree calculation for the observed distributions.

TABLE V. MEAN, MEDIAN, SKEWNESS, KURTOSIS, AND NUMBER OF SAMPLES FOR THE ALGORITHMS.

| Algorithm | Mean | Median | Skewness | Kurtosis | Samples |
|---|---|---|---|---|---|
| Hedera | 48.11 | 39.95 | 0.55 | -0.97 | 640385 |
| ECMP | 48.96 | 40.80 | 0.55 | -0.97 | 678761 |
| DCTCP | 51.14 | 43.35 | 0.49 | -1.09 | 770154 |

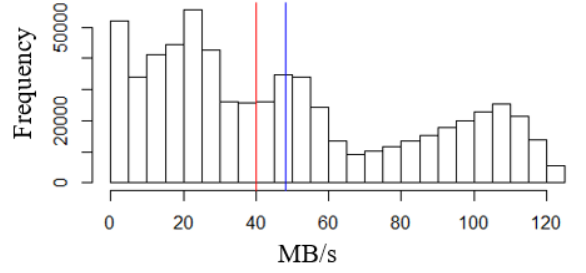

Fig. 4. Histogram of Hedera for the blocked rate.

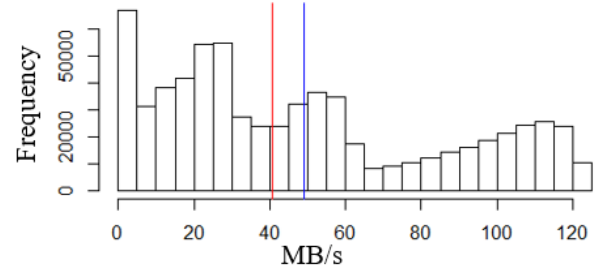

Fig. 5. Histogram of ECMP the blocked rate.

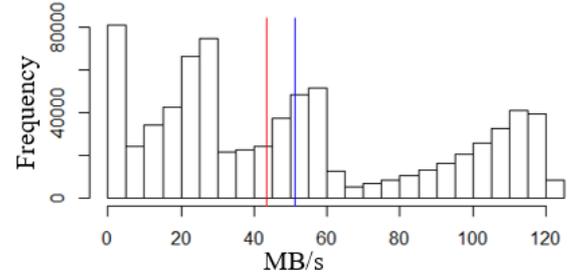

Fig. 6. Histogram of DCTCP for the blocked rate.

$$skew = \frac{\frac{1}{n}\sum_{i=1}^{n}(x_i - \bar{x})^3}{\left(\frac{1}{n}\sum_{i=1}^{n}(x_i - \bar{x})^2\right)^{\frac{3}{2}}} \quad (9)$$

Where $x_i$ holds $n$ observations and $\bar{x}$ is the mean values of the observations.

Kurtosis, as shown in equation 10, is the fourth central moment, and it is another essential shape measurement utilized for describing the distribution tail thickness compared to the Normal distribution. Typically, there are three types of Kurtosis, which are mesokurtic, leptokurtic, and platykurtic distributions. Mesokurtic distribution has the same characteristics of the Normal distribution concerning the extreme tail values, while leptokurtic has higher tail values due to the long tail, as for the platykurtic type, it has a precise tail with fewer outliers [31].

$$Kurtosis = \frac{\frac{1}{n}\sum_{m=1}^{n}(B_m - \overline{B_m})^4}{\left(\frac{1}{n}\sum_{m=1}^{n}(B_m - \overline{B_m})^2\right)^2} - 3 \quad (10)$$

Where $B_m$ holds $n$ observations of the predicted blocked rate and $\overline{B_m}$ is the mean values of the observations.

The calculated values indicate that all of the algorithms follow positive and semi-identical symmetry, but they are right-skewed since the mean values precede the median values. However, the degree of the skewness shows that the skews are moderate, which are between 0.5 and 1 [32]. In this case, the right-hand tail of the histograms will be longer than the left-hand tail, which means most of the data will be on the left-hand tail. But the length of the tail may affect the considering of the average value as the expected value of the loss rate [30]. However, we obtained kurtosis degrees for each algorithm to identify which one has the propensity to produce more outlier results. We found that the prediction distribution for the algorithms follows the platykurtic distribution since the kurtosis is negative compared with the Normal distribution. Therefore, the expected behavior for the algorithms is to produce fewer extreme values for the outliers at their tails, but it is clear that Hedera and ECMP have a higher degree of kurtosis, i.e., 0.79, in comparison with DCTCP what makes their highest loss rate not so trusted. Back to the histograms of ECMP and DCTCP in Figures 5 and 6, we noticed that the error rate to present its lowest values in the rage of 65 – 75 MB/s. However, these centrally located values may happen due to the throughput outliers' effects achieved from scenario 1:1 and 1:2 for both algorithms (Figure 3).

*D. Value at Risk (VaR) analysis*

Even though, the histogram and the statistics provide comparative information about the behavior of the model and the loss rate prediction, Value at Risk (VaR) analysis could provide more deep analysis based on some confidence [33]. The Monte Carlo simulation model considered as one of the three common types of VaR. In this research and for better generalizability, the chosen confidence level was 95%, since outlier results would appear with a more significant percentage, especially for Hedera and ECMP. Note that we calculated the probability of the confidence level by considering the quantile function, as in equation 11 [33].

$$VaR = -\mu_n + \emptyset^{-1}(1-u)\sigma_n \qquad (11)$$

Where $\mu_n$ is the mean of the values of the prediction, $\emptyset$ is the function of the standard Normal distribution, $\sigma_n$ is the standard deviation of the values and $(1-u)$ used for the chosen confidence level.

This kind of investigation presents a dynamic interpretation of how the elephant flows will be handled while employing such flow scheduling or congestion control algorithms. However, we depicted a broad examination of various confidence levels for the analysis in Figure 7. The loss rate in the case of Hedera is lowest with 112 MB/s for the total number of tested elephant flows appeared in Table 3. The loss of the others, i.e., ECMP and DCTCP, were 116 and 117 MB/s, respectively. Mainly, these values represent the maximum value that will be under risk of losing.

*5. The probability distribution of the whole overload*

In this section, we present the probability distribution of the entire workload, i.e., 120 connections of each scenario for all algorithms. We evaluate the performance of DCTCP, ECMP, and Hedera in terms of throughput of elephant flows and flow completion time of mice flows. All figures show the fact that Hedera and ECMP have very similar performance regarding flow completion time of mice flows and the throughput of elephant flows. Where Hedera employs ECMP for forwarding the mice flows, and ECMP performs well when there are no collide on switch ports what makes its performance in terms of elephant flows throughput closely approaches that of Hedera as shown in Figure 8(a), 8(b), 8(c). On the other hand, figures 9(a), 9(b), 9(c) depict the performance of DCTCP where its FCT of mice flows is more significant than that of Hedera and ECMP because of that DCTCP employs shallow threshold to trigger the marking event. Consequently, the transmission rate will be mitigated by sources where mice flow is delay-sensitive traffic, as well as elephant flows, have worse throughput than that of ECMP and Hedera where DCTCP provides flow control mechanisms, but it does not provide scheduling technique.

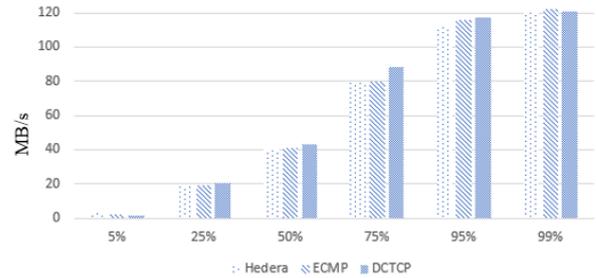

Fig. 7 Different confidence levels of VaR analysis.

In a nutshell, Hedera achieved a lower loss rate than ECMP as expected, but with higher variance for the error factor. We can infer that this factor makes the Hedera does not much outperform over ECMP. As for the response time, Hedera and ECMP achieved better flow completion time due to the static hashing between every source and destination on the network. In the case of flow congestion control in DCTCP, it has achieved its best in the 2:1 scenario whereas it has many outlier results in the 1:1 scenario as depicted in Figure 3. This indicates that the algorithm suffers in case of high elephant flow loads. Regarding data center applications that demand high bandwidth and low latency, every TCP loss causes bursty retransmission and that what makes queues length of the data center switches bloat frequently. Therefore, applications like MapReduce cannot make incremental progress without limiting the number of contending flows.

Therefore, we suggest that some fairness should be considered by providing a balance between link utilization,

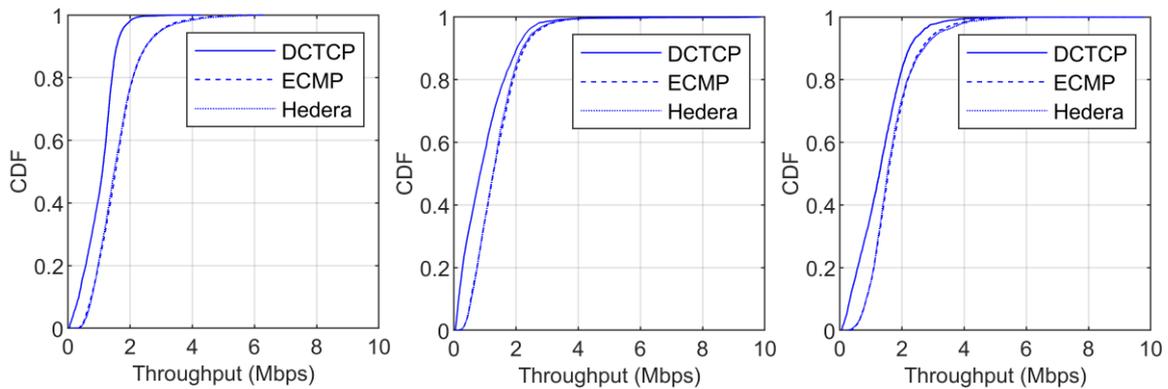

Fig. a Scenario 1:1  Fig. b Scenario 1:2  Fig. c Scenario 2:1

Fig. 8 Throughput of elephant flows.

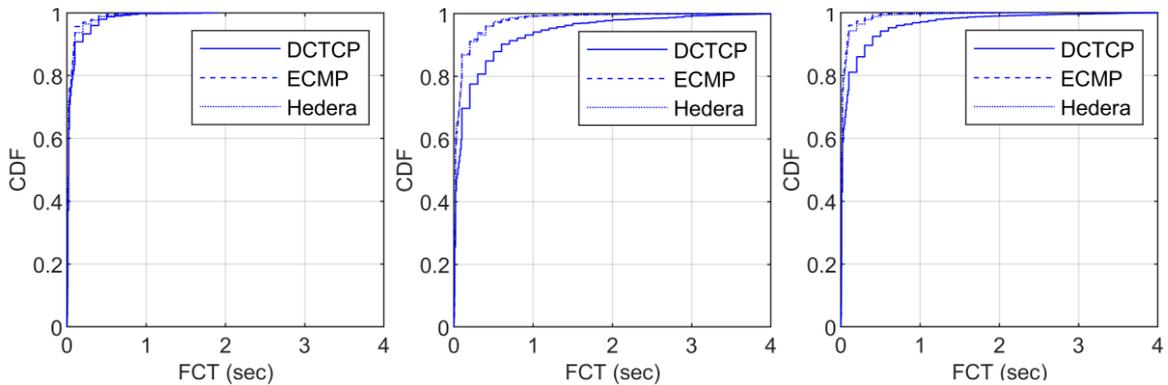

Fig. a Scenario 1:1  Fig. b Scenario 1:2  Fig. c Scenario 2:1

Fig. 9 Flow Completion Time.

congestion control. As for the performance evaluation methods of new algorithms that handle traffic flows, we recommend considering the uncertainty behaviors of the tested network and predict their loss rates. To the best of our knowledge, most of the developed heuristic algorithms for flow scheduling are evaluated using the average values for the obtained data without employing the probability distribution function. Note that the expected value for random variables does not exist for some distributions that have a long tail [30]. Consequently, considering the average for any sample of the data may not actually describe the expected value of the measured data, especially if the number of samples is limited. Such as in the case of Hedera [12] and Mahout [15] where the average value is taken for different performance evaluation objectives without identifying the proper probability distribution. Nevertheless, the essence of the prediction produced by independent and random variables relies on current observations to predict future performance. Accordingly, the model and assumptions need to be accurate enough.

## V. CONCLUSIONS

In this paper, we empirically designed, implemented, and analyzed a new performance evaluation model for flow scheduling and flow congestion control algorithms used in data center networks based on multiple stochastic workloads to predict the value at risk of the elephant flows loss rate. The evaluation considers the proper probability distribution functions for the proposed risk factors of the loss rate for Hedera, ECMP, and DCTCP. The proposed evaluation model has been built based on Monte Carlo simulation as a value at risk analysis model. The evaluation included an estimation of the probability distribution for risk factors based on Kolmogorov Smirnov and Anderson-Darling tests. Finding the probability distribution of such algorithms helps further mathematical analysis regarding elephant flow handling without conducting more practical experiments. The results of Hedera show that 64% of the evaluated TCP elephant flows are exhibited to be lost 112 MB/s with 95% of the confidence level, while ECMP lost 67.8% with 116 MB/s at risk, and DCTCP lost 77% with 117 MB/s. However, the throughput achieved by Hedera is not permanent due to the stochastic behavior of the traffic congestion. These risks have a direct influence on the status of data center applications in terms of flow completion time and throughput. However, the development of the flow scheduling techniques needs to have proper awareness in terms of flow risk analysis instead of accepting the simple average values of the results, especially when the samples are not large enough. Finally, further study is needed to evaluate more complicated data center workloads with real traces from data center applications to analyze more complex bottlenecks cases.

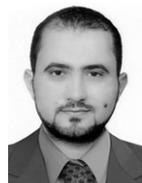

**Aymen Alawadi** Obtained his B. Sc. in 2006-2007 in the Computer Engineering from the University of Technology in Baghdad - Iraq. In 2007, he joined the University of Kufa in Najaf -Iraq as IT Engineer. From 2010 to 2012 he got M.Sc. in Computer Science from Universiti Sains Malaysia. Since 2017, he is a PhD student in the Department of Telecommunication and Media Informatics, Budapest University of Technology and Economics, Hungary.

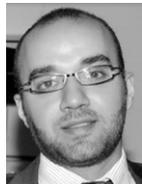

**Maiass Zaher** Obtained his B. Sc. in 2008 in the Computer Networks and Systems Engineering from Tishreen University, Syria. In 2013-2016 he got M.Sc. in Information Technology Engineering, Damascus University, Syria. Since 2016, he is a Ph.D. student in the Department of Telecommunication and Media Informatics, Budapest University of Technology and Economics, Hungary.

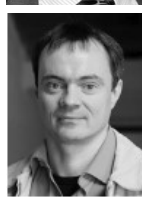

**Sándor Molnár** received his MSc, PhD and Habilitation in Electrical Engineering and Computer Science from the Budapest University of Technology and Economics (BME), Budapest, Hungary, in 1991, 1996 and 2013, respectively. In 1995 he joined the Department of Telecommunications and Media Informatics, BME. He is now an Associate Professor and the principal investigator of the tele traffic research program of the High-Speed Networks Laboratory.